# Laser Nano-Filament Explosion for Enabling Open-Grating Sensing in Optical Fibre

Keivan Mahmoud Aghdami[1,2], Abdullah Rahnama[1], Erden Ertorer[1,3], Peter R. Herman[1]

[1] Department of Electrical and Computer Engineering, University of Toronto, 10 King's College Rd., Toronto ON, M5S 3G4, Canada
[2] Department of Physics, Payame Noor University (PNU), P.O. Box: 19395-4697, Tehran, Iran
[3] Department of Physics and Pre-Engineering Program, Canisius College, Buffalo, NY, 14208, USA

## Abstract

Embedding strong photonic stopbands into traditional optical fibre that can directly access and sense the outside environment is challenging, relying on tedious nano-processing steps that result in fragile thinned fibre. Ultrashort-pulsed laser filaments have recently provided a non-contact means of opening high-aspect ratio nano-holes inside of bulk transparent glasses. This method has been extended here to optical fibre, resulting in high density arrays of laser filamented holes penetrating transversely through the silica cladding and guiding core to provide high refractive index contrast Bragg gratings in the telecommunication band. The point-by-point fabrication was combined with post-chemical etching to engineer strong photonic stopbands directly inside of the compact and flexible fibre. Fibre Bragg gratings with sharply resolved π-shifts are presented for high resolution refractive index sensing from $n_H = 1$ to 1.67 as the nano-holes were readily wetted and filled with various solvents and oils through an intact fibre cladding.

## Introduction

The compact and flexible format of optical fibre serves broadly as a high capacity conduit in today's information highway. These advantages further facilitate optical sensing of the local environment that can reach over short to long distances and probe into challenging environments [1,2]. By the nature of strong optical confinement, standard fibres require micro- or nano-engineering of devices directly inside of the core waveguide that is frequently applied by means of chemical etching [3], mechanical polishing [4], thermal tapering [5,6], laser modification [7,8], or ion milling [9]. Such structures redirect the waveguiding light to probe outside of the fibre cladding. However, optical probing is significantly more responsive when the external environment can be brought close to or directly into the fibre core as provided by photonic bandgap hollow-core fibre. The open structure enables refractive index (RI) sensing of gases [10] and liquids [11,12] but without the advantages for remote, localized or distributed sensing that is otherwise possible when micro-devices such as open cavities, micro-optical resonators and interferometers [13] have been embedded inside of a solid core fibre. Such micro-devices enable localized points of optical sensing at the core waveguide by basic means of absorption, fluorescence, scattering (Rayleigh, Mie, and Brillouin), diffraction, and interferometry [14-18]. These internal elements improve optical sensitivity that enable broadly-based lab-in-fibre applications in biology [19] or healthcare [20] such as label-free detection of cancer biomarkers [21].

Fibre Bragg gratings (FBG) are one favoured device in localized optical fibre sensing owing to sharp and environmentally responsive resonances [3]. The FBG further benefits from the ease of external laser fabrication through the cladding core [22]. However, the FBG typically probes the local environment through the cladding, without benefiting from the open photonic bandgap structure such as available in silicon photonic or other planar optical circuit technologies [17,23,24,25,26]. An open periodic structure with high RI contrast offers strong photonic bandgap responses which can be spectrally and spatially narrowed by optical defects to provide high-Q micro-resonators for highly localized sensing. Such micro-resonators enable optical trapping [26], label-free sensing [27] or single nano-particle detection [28] on a chip, but have been limited in fibre by a solid core FBG buried deeply in cladding. Various means of chemical, laser, and mechanical machining or thinning of the cladding have facilitated evanescent sensing at the surface of the FBG [14,29,30,31]. An open-structured photonic crystal has otherwise been challenging to fabricate transversely into the fibre core. Ion-milling or laser machining at the core of a thinned fibre has provided FBG sensors with nano-structured surface relief [32,33], blind holes [34] and through-holes [35,36]. However, the cladding processing renders such fibres mechanically fragile and lacking robustness for practical application. Yang et al. [37] accessed the core waveguide through a narrow cladding channel to open an array of micro-holes by chemical etching of self-focussed laser tracks. Third-order Bragg stopbands provided a microfluidic RI sensing response, but restricted to low sensitivity of 5 nm/RIU and narrow RI window from 1.32 to 1.41 owing to a large hole diameter (1.35 μm) exceeding the optical wavelength.

## Photonic bandgap engineering of the nano-hole Fibre Bragg grating

Femtosecond lasers provide an alternative direction for opening high-aspect ratio holes in transparent materials by a single-pulse nano-explosion from a filament-shaped beam focus. Hole diameters of ~200 nm diameter have been demonstrated in glasses by Kerr lensing, and axicon optics [38-40]. Our group has further harnessed surface aberration of glass plates to form similarly long filament tracks in bulk glass [41]. This approach was extended into optical fibre by using RI matching fluid to eliminate astigmatism of the cylindrical fibre shape and generate filament arrays having first-order Bragg stopbands [42-44]. With further scaling up of the interaction, we report here on driving a controlled filament explosion in the fibre cross-section, resulting in densely packed arrays of uniform nano-holes [45]. The isolated, blind or through holes were patterned with controllable positioning to selectively pierce the core and/or cladding and enabled strong photonic stopbands to be formed directly in the core waveguide with minimal processing of the surrounding cladding.

A schematic representation of the nano-hole array formation by laser filament nano-explosion is depicted in Fig. 1. The approach meets four key challenges for nano-structuring of strong photonic stopbands in optical fibre. Firstly, the beam delivery avoids astigmatic aberration by the cylindrical cladding shape (Fig. 1) resulting in formation of nano-holes partially to fully through the fibre cross-section, without thinning or inducing deleterious damage in the fibre cladding structure. Secondly, the nano-hole processing facilitates nano-hole assembly on small periodicity at optical wavelength scale, without inducing melting or significant heat-affected zone. The resulting 0.46 contrast in RI provides strong photonic bandgap responses without high optical scattering loss. Thirdly, the laser direct writing affords flexible and rapid patterning, to tune the spectral response and provide micro-cavity like responses such as π-shifted FBGs. Chemical etching facilitates photonic bandgap engineering by tuning the nano-hole diameter from 200 nm to 700 nm. And fourthly, the nano-holes draw significant capillary force to enable wetting by a wide range of solvent types fully through the 125 μm diameter of the fibre (Supplementary Video). The nano-hole array thus defines an open and flexible FBG sensing structure that can be fabricated rapidly in a single step procedure, and provide robust mechanical integrity together with strong opto-fluidic responses due to the sub-wavelength hole diameter.

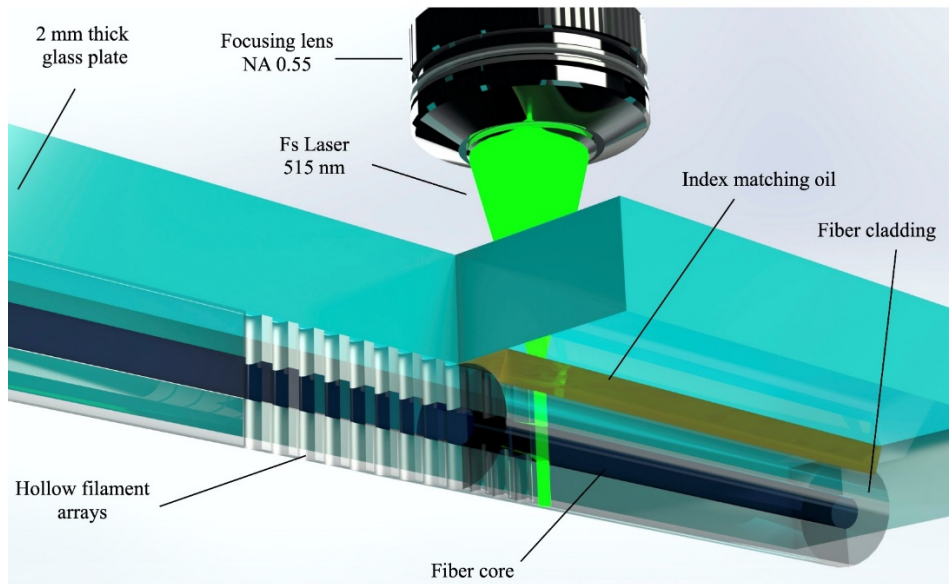

Fig. 1. Schematical arrangement of optical aberration plate and optical fibre used for generating femtosecond laser filaments, and opening high aspect ratio nano-holes through the fibre cladding and core waveguide.

The anticipated optofluidic responses of the nano-hole FBG structure (Fig. 1) are significantly stronger and more sensitive than with the traditional FBG grating as demonstrated by the simulations in Fig. 2. The second-order grating responses were modelled in standard telecommunication fibre (SMF-28) having 300 nm diameter holes with 1072 nm periodicity and relatively short length of 643 μm. The potential for precise refractive index (RI) sensing is noted by the strong and rapidly shifting stopbands (Fig. 2a) when nano-holes are filled with air and solvents having different RI values spanning from $n_H$ = 1 to 1.66 (see also Fig. S1). The sub-wavelength hole diameter provides increasingly stronger (i.e., > -2 dB for air) and broader resonances (up to 7 nm bandwidth), rising up for both increasing or decreasing refractive index from the matching condition at $n_H \cong n_{eff}$ = 1.450 (Fig. 2a). With variable hole diameter, the EME modelling indicated a high RI sensitivity response of up to 600 nm/RIU will be available at high RI (i.e., $n_H$ = 1.66) and large hole diameter of 700 nm.

When imposing a π-shifted defect into the nano-hole array, the simulated stopband (Fig. 2a, inset spectrum) opens a sharp and narrow transmission resonance (Δλ = 100 pm, 3 dB bandwidth) inside of the air-filled stopband, representing a moderately strong resonator quality factor (Q ~ $10^4$). The strong influence of the π-defect and nano-hole array on the guided light field distribution is demonstrated by the high repulsion (Fig. 2b) or attraction (Fig. 2c) of the mode within a narrow (~2.5 μm) zone following along nano-holes filled with air ($n_H$ = 1) or oil ($n_H$ = 1.66), respectively. For the higher RI case ($n_H$ = 1.66), magnified views of the intensity profiles near the π-defect (x = 0) reveal an asymmetric narrowing from the fundamental mode field from 10.4 ±0.5 μm diameter to 5 μm (Fig. 2d) and 8 μm (Fig. 2e) for respective perpendicular and parallel directions with respect to the hole axis. The beginning of a micro-cavity response is noted along the fibre centre axis (Fig. 2f) by the light intensity drawing into a short 66 μm zone (full 3 dB) as presented for the air ($n_H$ = 1) and oil ($n_H$ = 1.66) cases (Fig. 2f).

## Results

### Nano-hole array Fibre Bragg Grating.

Long and uniform laser filament shapes (Fig. 1) of up to 125 μm length were provided by surface aberration from glass plates of up to 3 mm thickness. In this arrangement, high laser pulse energies of 7 μJ drove a uniform nano-filament explosion without distortion from Kerr lens focusing or plasma defocusing effects. The nonlinear optical benefits of narrowing modification dimensions below diffraction limited sizes was thus retained. Isolated, blind and through holes could be tailored by the laser and focussing controls to any position in the fibre cross-section. An example of an FBG with nano-hole arrays reaching fully through the cladding and core is presented in Fig. 3. A schematic of the FBG (centre image) represents the dense packing of the nano-holes on 1.072 μm periodicity, targeting a high aspect ratio of processing control reaching through the 125 nm diameter fibre cladding. The evidence for this high-aspect ratio hole geometry is provided in the time sequence of microscope images showing the evaporation of isopropanol from fully wetted (i.e., air filled) nano-holes, recorded over ~10 seconds (see also supplementary video). A wide range of solvent types were found to readily wet and fill the nano-holes of such laser structured fibre.

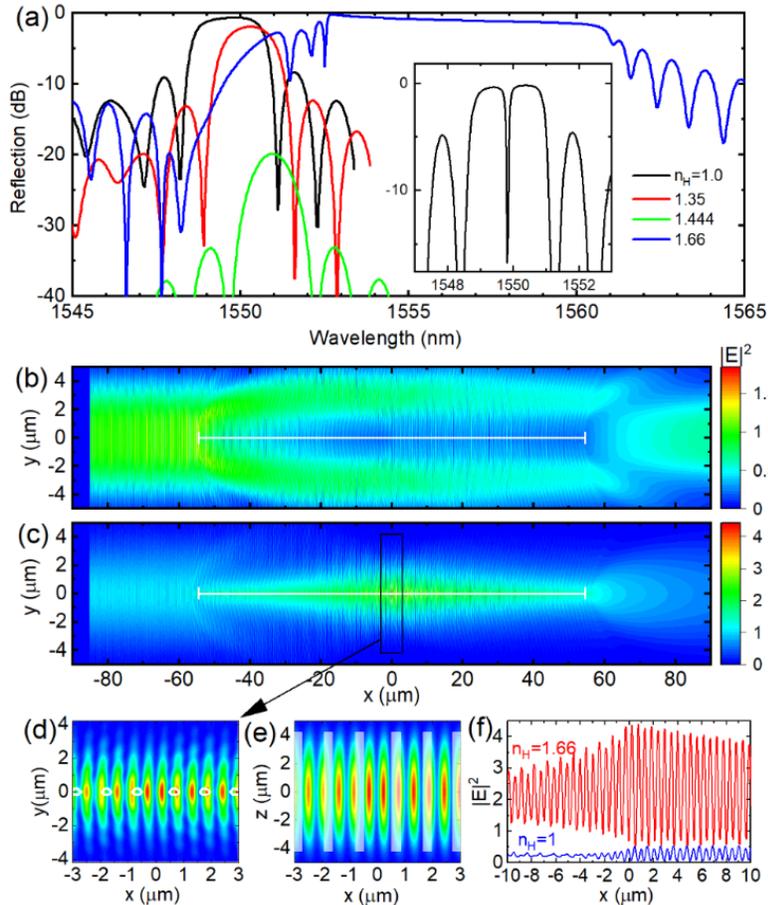

Fig. 2. Simulated reflection spectra (a) for a second-order FBG of uniformly pitched nano-holes (Λ = 1.072 μm) having 300 nm diameter, 0.643 mm array length (600 holes) and high RI contrast (Δn = -0.45 to +0.21) when filled with air, solvent or oil. A sharply resolved transmission band (30 pm FWHM) has opened (inset spectrum, $n_H$ = 1) from a π-shifted FBG of 1200 holes. The simulated intensity distribution transverse to nano-holes (xy plane) for low ($n_H$ = 1.0) (b) and high ($n_H$ = 1.66) (c) RI cases of a π-shifted FBG (100 holes, x = -53.6 to +53.6 μm). Magnified views for the high RI case reveal intensity concentration in the filament plane (e) and around the π-defect (d; x = 0). White circles and bars outline the hole positions. Axial intensity oscillation (x axis, y = z = 0) comparing (f) high and low RI cases.

The provision of refractive index matching oil around the fibre cladding (Fig. 1) did not inhibit near-neighbour filament nor nano-hole formation. Lateral damage zones were insignificant both at surfaces and the internal volume such that isolated holes could be formed, pulse by pulse, in arrays having periods as small as 1.072 μm (Fig. 3f, g). The point-by-point writing permitted insertion of optical defects such as the π-shift identified in the optical microscope image (Fig. 3f).The end-view of the nano-channel array, formed on 1.072 μm periodicity confirmed a sub-micron lateral resolution of hole positions inside the core waveguide, and extending well beyond the beam depth of focus (1.4 μm) to reach through the fibre cladding cross-section (Fig. 3g).

The cross-sectional morphology of the nano-hole structures was further examined by SEM imaging. A section of nano-hole FBG was cleaved and polished at an oblique angle (~30°), as shown schematically in Fig. 3a. Electron microscopy of the oblique fibre cross-section identifies an array of nano-holes having formed through the core and cladding (Fig. 3b). With increasing magnification of the core section, the sequence of SEM images (Fig. 3b-e) confirm the formation of isolated nano-holes having relatively uniform cylindrical shape, and positioned on $\Lambda = 1.072$ μm period with an estimated hole diameter of ~200 nm. A slightly irregular hole perimeter is noted with an internal surface roughness of ~50 nm, which is ~30 × smaller than the optical probe wavelength. Hence, the holes can form fully isolated in tightly packed arrays having nearly uniform cross-section through the core (Fig. 3) and a majority of the cladding. The nano-hole arrays are only noted to break into each at the cladding surface over a depth of ~15 μm. Hence, nano-holes with high aspect ratio greater than 500× have been demonstrated in the fibre.

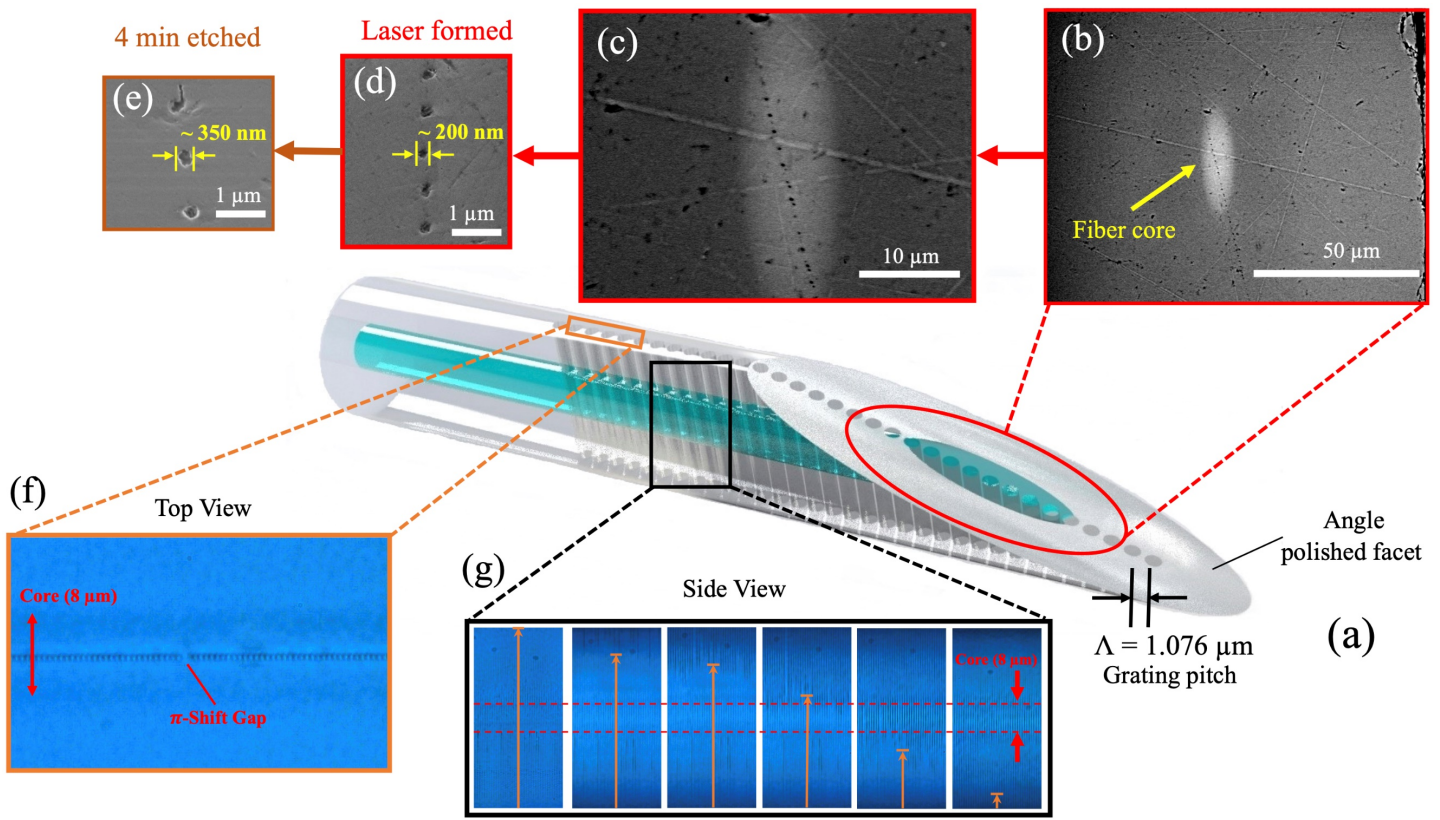

Fig. 3. Schematical view of a nano-hole embedded optical fibre with angled facet (a) and supporting microscopic imaging (b-g). Optical microscopy of the nano-hole array showing central alignment of the array to the core in the end view (f) and a time sequence (left to right) of evaporating isopropanol with menisci level (yellow arrow) moving downward in side views of the nano-holes (g). Under increasing magnification, SEM images of the angled fibre facet (a-d) unveil open nano-holes having formed continuously and uniformly (~200 nm diameter) through the core (white oval zone) and cladding without breakthrough on the tight 1.072 μm periodic spacing. SEM (e) shows the hole diameter expanded to 350 nm after 4 min of chemical etching (5% HF).

Emersion of the FGBs in diluted hydrofluoric acid (5%) provided a second processing step to reproducibly increase the hole diameter, for example, to ~350 nm diameter after 4 minutes (Fig. 3e). Such diameter tuning opens a means for tuning the photonic stop band response, as well as increasing the FBG sensitivity to refractive index changes.

Photonic stop band response.

The validation of strong and responsive photonic stop bands is presented in the reflection spectra of Fig. 4a for nano-hole arrays filled with air and a range of solvents (Table 1). Under high refractive index contrast with air ($n_H$ = 1.0) or Oil10 ($n_H$ = 1.66), the nano-hole array provided broad (~2 nm) and strong (~3 dB) Bragg resonances with only 600 holes (i.e. 643 µm length). The evolution of the spectral profile, reflection peak and linewidth with increasing grating length for the case of air-filled holes (Fig. S2) followed the growth trends of traditional FBGs, expect developing much more rapidly due to a nearly 100-fold higher refractive index contrast ($\Delta n$ = 0.45) over traditional FBGs ($\Delta n \sim 10^{-3}$ [2]). Although the 200 nm hole diameter encompassed only a ~2% overlap with the modal field (MFD ≅ 10.4 µm), the gratings provided a strong effective coupling strength of up to $\kappa_{AC}$ = 9.17 cm$^{-1}$. The strong responses are enabled by the strong field repulsion or attraction effects of the guided light into or out of the planar grating zone (Fig. 2b versus Fig. 2c) for the cases of low and high refractive index, respectively. As a result, the Bragg resonance had shifted definitively from 1549.93 nm in air ($n_H$ = 1.0) to 1551.76 nm in Oil10 ($n_H$ = 1.661), providing a average RI sensitivity of 2.75 nm/RIU across a wide refractive index sensing range. The progression of strong to weak stop bands, from -2.72 dB reflection with air to -17.28 dB with Oil5 ($n_H$ = 1.448), marks the matching condition on refractive index, beyond which higher refractive index (i.e. Oil10, $n_H$ = 1.661, Fig. 4a) regenerated strong stopbands (-2.76 dB).

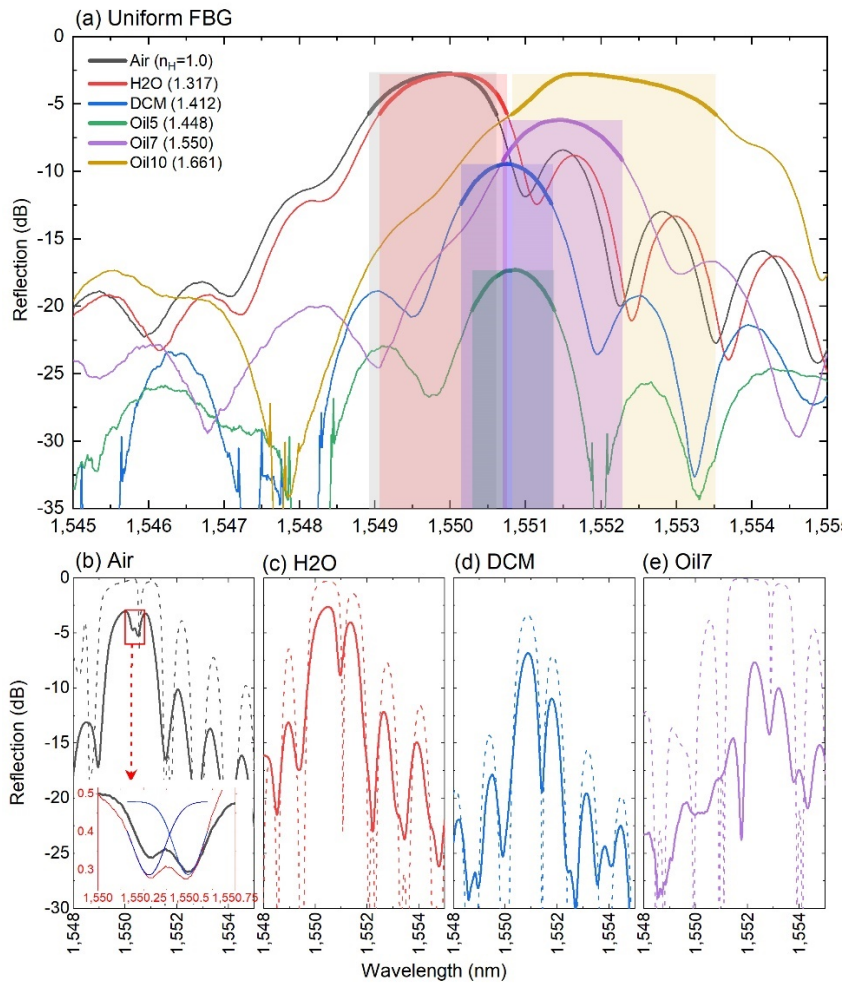

Fig. 4. Reflection spectra (a) recorded from a uniform FBG of 600 nano-holes ($E_{pulse}$ = 4.5 µJ, $\Lambda$ = 1072 nm, 2 mm plate) forming strong (3 dB) and wide (~2 nm) stopbands when filled with low ($n_H$ = 1) or high ($n_H$ = 1.661) refractive index materials. The coloured bands mark the 3-dB bandwidth. Refractive index matching oil ($n_H$ = 1.448) dramatically weakens the stopband. Reflection spectra (solid lines) of unpolarized light (b)-(e) recorded from a π-shifted FBG of 1,200 nano-holes for four different examples of RI values ($n_H$ by colour from legend (a)) and comparison with simulated spectra for P-polarized light (dashed lines). Laser induced birefringence of $\Delta\lambda$ = 210 pm is resolved for the air-filled case ((b) inset). Gaussian line fits of the π-defects (blue and red lines in (b) inset) provide ±10 pm spectral precision.

Although strong in reflection, the breadth of the stopbands under high refractive index contrast are limiting in specifying the precise centre Bragg wavelength. A more rewarding direction was obtained by implanting a π-defect into a Bragg grating array of 1200 nano-holes, opening well resolved pass-bands (Fig. 4b-e, Fig. S6). In the progression for nano-holes filled with air, water, Dichloromethane and Oil7 ($n_H$ = 1.0, 1.316, 1.412 and 1.550 in Fig. 4b, c, d, and e, respectively), the π-defect provided the highest contrast (> 10 dB) and sharpest resolution ($\delta\lambda_B$ = 150 pm) pass-bandwidth (3 dB) when tuned nearer to the refracting index matching condition, $n_H \cong n_G$ (Fig. 4d and see also Fig. S6a). For air-filled nano-holes, the resonance separated into two ~3 dB peaks ($\delta\lambda_B$ = 210 pm, Fig. 4b), unveiling a waveguide birefringence of $\Delta n_H$ = 0.07. Otherwise, the birefringence was unresolved and served only to broaden the pass band ($\delta\lambda_B$ = 150 pm to 400 pm) with increasing refractive index contrast examined in range 1.31 < $n_H$ < 1.67 (Fig. S7a). The point-by-point writing of such high aspect ratio nano-holes with <100 nm positional precision thus facilitated photonic band shaping with sharp spectral features tuned to ~100 pm resolution.

EME modelling (Fig. 4b-e, dashed line spectra) provided a good representation of the π-shifted grating reflection (Fig. 4b-d, solid line spectra) with the hole diameter optimized to ~200 ± 10 nm for the 1200 nano-hole array design ($\Lambda$ = 1.072 μm). The simulated spectra have reproduced the broad stopband and sideband features as well as the narrow π-defect response, and accurately tracked the wavelength shifts for the full range of refractive index changes from $n_H$ = 1.0 (Fig. 4b) to 1.550 (Fig. 4e). However, the grating reflection fell short of the simulated peak values by 3 dB to 7 dB over the respective air ($n_H$ = 1.0) to Oil7 ($n_H$ = 1.550) cases, pointing to unaccounted losses from Rayleigh scattering on nano-hole surface roughness (~50 nm, Fig. 3), variances of up to ±50 nm on the nano-hole positioning, and first order Bragg radiation.

To improve on the resolution for RI sensing, Gaussian shaped functions were found to best match the π-defect transmission dip, for example, as shown for the case of the air-filled holes (Fig. 4b inset). The spectral matching enabled centre resonant wavelengths to be specified with a precision of ±10 pm. Otherwise, the Gaussian representations furnished linewidths varying from 180 to 360 pm (FWHM) for the unresolved birefringence cases (Fig. 4c-e; see also Fig. S6).

## Photonic-bandgap engineering – etching.

The as-formed nano-holes provided added utility in guiding chemical etchant to tune the hole diameters and further engineer the stop band response. SEM revealed a near-linear response of increasing hole diameter with etching time, beginning from the ~200 nm diameter for the laser-formed nano-hole and expanding to 350 nm for the case of 4 min etching in 5% HF acid (Fig. 3e). In real-time monitoring of FBG reflection spectra (1200 holes), the stopbands in both uniform and π-shifted gratings (Fig. S3a supplemental) shifted monotonically to shorter wavelength over a 7 min etching time. The blue spectral shifting arises from a drop in effective refractive index as the lower refractive index of etchant ($n_H \cong n_{H2O} \cong 1.3164$) displaces the higher refractive index glass ($n_G$ = 1.45). An increasing hole diameter is noted to weaken and narrow the stopbands (Fig. S3b) from -3.7 to -29 dB in strength and 1.5 to 0.97 nm in bandwidth, over the 7 min etching time. However, the π-shifted passband (not plotted) retained a full 9 dB contrast inside of the FBG stopband, even as the stopband weakened by > 20 dB over the full 7-min etching time (Fig. S3 b). The preservation of sharply resolved π-defect resonances attests to a highly ordered and unbroken patterning of the high aspect ratio nano-holes, that do not break through on a tight packing density (1.072 μm period) even as diameters were opened up over the 7 min etching time.

Nano-hole FBGs with uniform gratings were prepared with different etching times (0, 2, 4, and 6 min) to provide a widely varying FBG response (Fig. S4) when filled with solvents or oils spanning a large range of refractive index values (Methods Table 1). The nano-hole arrays as formed by the laser (0 min) typically offered the strongest stopbands (Fig. 5a). The progression to weaker stop bands (Fig. 5a-c) varied from strongly to weakly for solvents having the lowest (Fig. 5a, $n_H$ = 1), matched (Fig. 5b, $n_H$ = 1.448) and highest (Fig. 5c, $n_H$ = 1.6) values of refractive index. This progression is noted in the plot of peak Bragg reflectance (Fig. 5d) over the full range of tested refractive index values (Table 1). The fall off was most pronounced for air ($n_H$ = 1), decreasing by greater than 25 dB over the 6 min etching time. The fall off was delayed in the case of higher refractive index oils with refractive index values of $n_H$ = 1.522 to 1.66, where reflectivity first increased to a maximum of ~5 dB for 2 to 4 min etching time. In all cases, the stopbands were weakest for the 6 min etching time. The bandwidth (3dB) of the Bragg stopbands were also strongly influenced by the chemical etching time, either narrowing by ~50% or broadening by more than 3-fold according to the negative ($n_H$ < 1.45) or positive ($n_H$ > 1.45) contrast of refractive index (Fig. 5e).

The alignment of the EME modelled spectra (Fig. S1) with the observed reflection spectra (Fig. 4, S4) provided a precise, semi-empirical estimate of the nano-hole diameter, for example, yielding 500 ± 20 nm diameter for 4 min etching time (Fig. S5). In this way, hole diameters of 220, 300, 500, and 700 nm were assigned with variances of ~± 10 nm to FBGs opened with 0, 2, 4, and 6 min etching time, respectively. The spectral corroboration identifies a peak value of FBG reflectance (Fig. S3, 5d) arising on the first quarter wavelength resonance of the nano-hole diameter (i.e., $\lambda/4n_H$), for example, encompassing hole diameters of 220 to 300 nm for solvents varying from $n_H = 1$ for air to $n_H = 1.41$ for Dichloromethane. The steep fall-off of reflection thus aligns with an anti-resonance on doubling of the hole diameter to $\lambda/2n_H$, corresponding to diameters of 500 to 700 nm for air to high index oil ($n_H = 1$ to 1.60). Hence, the diameter of nano-holes transitions from a first order resonance for the strongest stop reflection band (0 - 2 min etching) to the first anti-resonance (6 min) over the ranges of laser formation and chemical etching tested here.

An increasing hole diameter further provided stronger spectral shifts of the stopbands (Fig. S4, 5a-c), moving in reversed directions as expected depending on the positive ($n_H > n_G$) or negative ($n_H < n_G$) contrast of solvent refractive index with respect to the glass index. The largest hole diameter (~700 nm for 6 min) and highest refractive index oils offered the highest wavelength sensitivity, with the Bragg centre wavelength shifting by up to +40 nm for $n_H = 1.60$ solvent over the 0 to 6 min etching time. In contrast, a smaller and negative wavelength shift of -2.2 nm was noted (Fig. 5f) for the case of the air-filled holes over the same 6 min etching time. Under index matching (Fig. 5c, $n_H = 1.45$), the Bragg resonance did not shift.

## High-resolution RI sensing

In order to provide the highest resolution RIU sensing, π-shifted FBGs of 1200 hole arrays were modified with similar etching times (0, 2, 4, and 6 min). EME modelling offered close matching of the spectra strong bands, side lobes, and π-shift passbands (Fig. 4b-e), yielding similar values of effective hole diameters, corresponding to 220, 300, 500, and 700 nm for the respective 0, 2, 4, and 6 min etching times. The π-defect pass bands (Fig. S6) were spectrally fitted to Gaussian line shapes, varying from 200 to 400 nm linewidth (Fig. S7). A narrow birefringent splitting (~200 pm) of the π-defect was occasionally resolved in the spectra (Fig. S7) in cases with the strongest stopbands and largest contrast in refractive index.

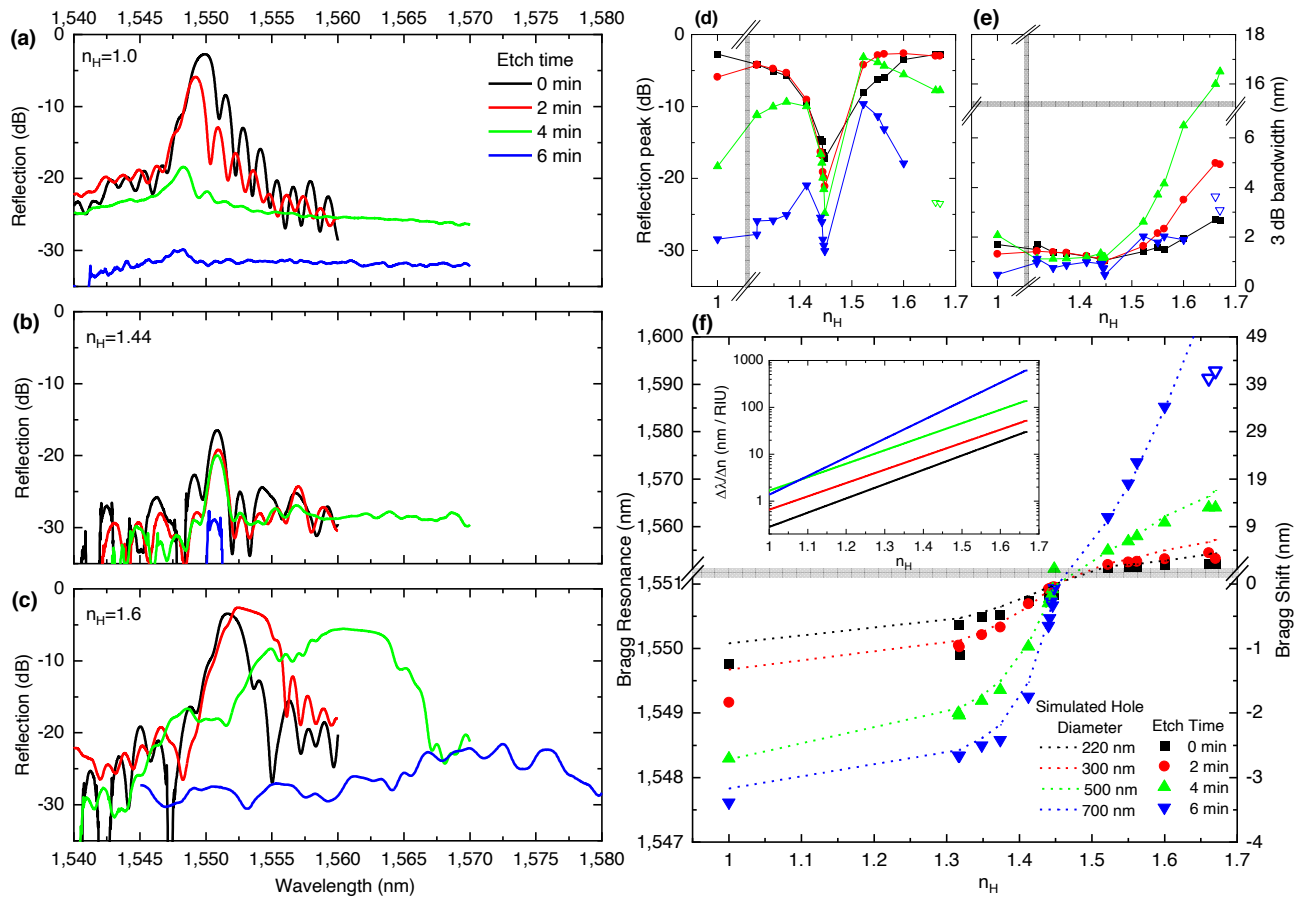

Fig. 5. Reflection spectra recorded from uniform FBGs ($E_{pulse}$ = 4.5 μJ, Λ = 1072 nm, 600 holes) filled with low ((a) $n_H = 1$), medium ((b); $n_H = 1.44$), and high ((c); $n_H = 1.661$) refractive index materials, revealing the influence of increasing nano-hole diameter due to varying chemical etching time (0 - 6 min). Peak reflection strength (d) and 3-dB linewidth (e) of the Bragg resonances plotted as a function of refractive index for different chemical etching times (following legend (a)). Bragg resonance wavelength (shifts on right axis) plotted versus refractive index (f) for different chemical etching times. A global fit of the data by EME simulation is presented (dashed lines) for hole diameters of 220 to 700 nm (see legend). The slopes of data provide a strongly increasing refractive index sensitivity (inset) for increasing RI and increasing hole diameter (colour coded lines). Shaded zones mark axis scale changes.

With spectral line fitting (Fig. S6) of the π-defects, centre Bragg wavelengths could be determined to ±10 pm precision, enabling RI sensing to a high resolution of $10^{-5}$ RIU. When plotted against refractive index (Fig. 5f, solid fonts), the π-resonance shifts demonstrated an impressive RI response of FBG stopbands, shown globally over an extraordinary range of refractive index values ($n_H$ = 1 to 1.66) and nano-hole diameters (220 to 700 nm). The EME modelling (Fig. 5f, dashed line) followed each data set to ±200 pm (rms) spectral precision, relying only on one value of optimized hole diameter across the full refractive index testing range. Discrepancies in air (±500 pm) and between methanol and water may arise from surface tensions effects that require further study. Sharply forming π-defect resonances were identified in all cases except the two highest indexes ($n_H$ = 1.661 and 1.670) and large diameter (6 min) condition (open triangle Fig. 5f), where stopbands became overly broad and mixed with side bands (see Fig. S4d).

The narrow π-shifted stop bands provided strong optical responses to sense the local environment, as demonstrated in the wide 1548 to 1590 nm shift in Bragg wavelength (Fig. 5f). The slopes of these responses provided a widely range RIU sensitivity response (Fig. 5f, inset) that reach as high as 600 nm/RIU for the case of highest refractive index ($n_H$ = 1.66) and largest hole diameter (700 nm). This RIU sensitivity is comparable to the best FBG-based demonstrations to date (i.e., 945 nm/RIU in [46]). Moreover, spectral line fitting (±10 pm) offered a high-resolution determination of the refractive index to ±$10^{-4}$ precision.

The nano-hole array presents a novel optical fibre sensor, retaining cladding strength while readily wetting with numerous solvents that can reach into the open photonic structure formed along the fibre core. The facile means of laser writing and patterning of high-aspect ratio nano-holes thus opens a new realm of creating strong and compact photonic-stop band devices directly in traditional optical fibres while facilitating environmental sensing through a thick robust cladding. The fabrication method is extensible to other types of fibres and fibre materials, including the two-dimensional patterning of nano-holes arrays. In this way, the minimally invasive methods of filament explosion and chemical etching are enabling in the photonic bandgap engineering of traditional optical fibre, promising to transform how fibres shape the flow of light and sense the local environment from applications in biomedical probes through to large area communication networks.

# Methods

## Laser fabrication of hollow-filament arrays.

A frequency-doubled Yb-doped fibre laser (Amplitude Systems, Satsuma) provided femtosecond pulses of 515 nm wavelength, 250 fs pulse duration and $M^2$ < 1.2 beam quality. The 800 kHz repetition rate was down counted to 1 Hz, to provide up to 8 µJ pulse energy to the filament forming exposure arrangement (Fig. 1). The laser beam was expanded ~2 times (Linos Magnification Beam Expander, 4401-257-000-20) to fill the full aperture of an aspherical focusing lens (New Focus, 5722-A-H) of 0.55 NA. A 0.8 µm spot radius (1/$e^2$ intensity) and 1.4 µm depth of the focus is expected if focussed directly into silica glass.

In the arrangement of Fig. 1, a long and uniform filament beam shape with diameter of < 1 µm was generated by spherical aberration induced by the flat surfaces of fused silica plates (Nikon, NIFS-S (S-grade)) positioned between the lens and optical fibre. Filament track lengths from 80 to 125 µm were generated with 2 and 3 mm plate thickness, respectively. The surface aberration stretched the laser energy over a large depth of focus in a similar way to an axicon forming a Bessel-like beam [39]. In this way, high pulse energy was applied without inducing distortion from Kerr lens focusing and plasma defocusing effects, while retaining the nonlinear interaction benefits of narrowing modification to sub-diffraction limited size.

A standard single mode telecommunication fibre (Corning, SMF-28) was mounted in contact with the bottom surface of the aberration plates (Fig. 1), following procedures previous developed for filament writing of low-contrast FBGs [42]. Index matching oil (Cargille, 50350) was applied to fill the gap between the fibre and plate and remove cylindrical aberration by the fibre. Laser filaments were aligned laterally to ±1 µm precision to bisect the waveguide core (Fig. 3f), and shifted vertically to illuminate any portion of core and cladding (Fig. 3g) by using a three-axis motion and alignment system (Aerotech Inc., Aerotech-PlanarDL-00 XY and ANT130-060-L Z) having 3 nm resolution. Filament positions in the fibre cross-section (Fig. 3f, g) were verified by backlighting of the fibre with an optical microscope (Olympus, BX51).

The laser pulse energy was adjusted upward from the low-contrast conditions until the optical contrast of filament tracks forming in the fibre was observed to darken and indicate the onset of micro-explosion, opening hollow filament shapes inside of the fibre. Single pulse exposure of 3 and 7 µJ energy opened hollow filaments of 80 µm to 125 µm length, respectively, with silica plates of 2 and 3 mm thickness, respectively. The axial focusing was optimized to avoid the burning or evaporating of the refractive index matching oil and minimize ablative machining and near-surface damage at the fibre cladding surface. In this way, blind (2 mm plate) or fully open (3 mm) holes (Fig. 3g) were formed in the fibre core and cladding. The filament tracks could be aligned side-by-side to ~1 µm spacing and remain isolated without breaking through or undergoing beam propagation or processing distortion by the pre-existing track. Filament tracks were assembled into a tightly packed linear array, and precisely centred along the fibre core (Fig. 3f, g) for evaluated for FBG responses.

### Nano-hole morphology.

The nano-scale formation of high aspect ratio holes into the optical fibre was examined by mechanically cleaving and optically polishing of a laser-exposed fibre at an oblique angle (~30°) (Fig. 3a). A filament array of nano-holes was generated through the cladding and core waveguide regions on 1.072 µm period, using 4.5 µJ pulse energy, and 2 mm thick aberration plate. The cleaved facet was polished with silicon carbide paper (1200 grit), while wetted with few drops of distilled water. After cleaning with acetone and isopropanol, the polished fibre facet (Fig. 3a) was viewed in orthogonal alignment with a scanning electron microscope (Hitachi, SU5000), providing high contrast views of the full fibre cross-section, shown in Fig. 3b-d. The scanning electron microscopy images confirmed a ~200 nm diameter of the open nano-holes forming through a majority of the filament length, defining a high aspect ratio of ~500 times. Hole-to-hole break through was not observed over most of the filament length for periods as small as $\Lambda = 1.072$ µm, or for identical nano-hole arrays opened to larger diameter (i.e., 350 nm, Fig. 3e) after chemical etching.

### Capillary flow.

Optical microscopy was used for confirming and monitoring the wetting, capillary flow and evaporation of liquids with various values of refractive index (Table 1) into and out of the laser formed nano-holes, providing optical images (Fig. 3g) and video recordings (supplementary video). The wetting times varied from sub-seconds for solvents (i.e., water, methanol, acetone, etc.) (Merck) to tens of seconds for index matching oils (Cargile). Dichloromethane (DCM) and acetone were applied to remove oils and to clean nano-holes before immersing the fibre in a new liquid.

Table 1: Refractive index (RI) of liquids applied inside of nano-hole FBGs

| Liquid | Abbreviation | RI at $\lambda = 1550$ nm [47] |
|---|---|---|
| Air | Air | 1.002 |
| Water | $H_2O$ | 1.3164 |
| Methanol | MeOH | 1.3174 |
| Acetone | Ace | 1.3483 |
| Isopropanol | IPA | 1.3737 |
| Dichloromethane | DCM | 1.4124 |
| Cargile AA-1.450 | Oil1 | 1.440 |
| Cargile AA-1.452 | Oil2 | 1.442 |
| Cargile AA-1.454 | Oil3 | 1.444 |
| Cargile AA-1.456 | Oil4 | 1.446 |
| Cargile AA-1.458 | Oil5 | 1.448 |
| Cargile A-1.540 | Oil6 | 1.522 |
| Cargile A-1.572 | Oil7 | 1.550 |
| Cargile A-1.586 | Oil8 | 1.562 |
| Cargile A-1.630 | Oil9 | 1.600 |
| Cargile B-1.700 | Oil10 | 1.661 |
| Cargile M-1.705 | Oil11 | 1.670 |

### Fibre Bragg grating characterization.

At low pulse energy exposure (~350 nJ), low-contrast filament arrays were previously shown to assemble in silica fibre on $\Lambda = 0.536$ µm period and provide strong first-order Bragg grating responses at 1550 nm [42]. However, the opening of nano-holes with the higher pulse energy (3 to 7 µJ) required in the present work resulted in filament distortion and break-through of holes when positioned on a similar first-order period. Formation of isolated nano-holes were verified by SEM (Fig. 3d) with a doubling of the period to $\Lambda = 1.072$ µm, thus enabling a second order FBG response at 1550 nm wavelength. π-shifted FBGs were fabricated for narrowing the spectral response of device to below 200 pm linewidth (3 dB). FBG spectral responses were recorded real-time during the laser fabrication and chemical etching, or during the filling or evaporation of various liquids (Table 1). Reflection spectra were excited with a 1530–1610 nm wavelength broadband source (Thorlabs, ASE-FL7002), and recorded through an optical fibre circulator (Thorlabs, 6015-3-FC) by a high-resolution optical spectrum analyser (Anritsu, MS9740B). The influence of laser exposure and number of filaments were evaluated. Fig. S2 shows the influence and limitation of increasing device length on the reflection peak and bandwidth. A 600 and 1200 element array of nano-hole filaments were selected to study the FBG responses to solvents, chemical etching, and to compare uniform and π-shift gratings.

### Chemical etching.

Femtosecond laser irradiation followed by chemical etching (FLICE) [48] was adopted here for opening the nano-hole diameter. FLICE extended the diameter of the laser-formed nano-holes in a predictable and reproducible way. The laser formed FBGs were immersed in acetone and followed with DCM to remove oils and debris. Fibres were submerged in 5% dilute hydrofluoric (HF) acid solution for up to seven minutes. Longer etching times were found to degrade the FBG spectrum. After etching, the fibre was promptly immersed in distilled water, then IPA and left to air dry. Identical FBGs were prepared with 2, 4, and 6 min of etching and then spectrally characterized under filling with all solvent and oil types listed in Table 1. Assessment by a combination of SEM morphology (Fig. 3e), FBG spectral responses (Fig. 5f and Fig. S4), and simulations (Fig. S1) provided an estimated etching rate on the nano-hole diameter beginning more slowly at 40 μm per minute in the first two minutes and rising to a steady value of 100 nm/minute thereafter.

### Optical modelling.

Simulation of the 3D light intensity distribution (Fig. 2b, c) and the spectral reflections (Fig. 2a and Fig. S1) expected from a nano-hole array positioned in the core waveguide of SMF-28 fibre was provided by commercial software (Lumerical Inc.), based on finite difference time domain (FDTD: 3D Electromagnetic Simulator) and eigenmode expansion methods (EME; MODE: waveguide simulator), respectively.

For the FDTD simulation, intensity patterns (Fig. 2b, c) were provided for a second-order FBG of 100 nano-holes on $\Lambda = 1.072$ μm pitch and having nano-hole diameter of 300 nm. Core and cladding refractive index values were matched to SMF-28 fibre at 1550 nm wavelength. A perfect electrical conductor was imposed at the centre-fibre symmetry plane (z = 0 in Fig. 2 b, c) with absorbing boundary conditions on the outside borders. A spatially uniform mesh with grid size of $\Delta x = \Delta y = \Delta z = 48$ nm was applied with time steps of $\Delta t = 0.092$ fs. Long program running time limited simulations to FBGs having a linear array of 100 filaments that spanned ~100 μm along the fibre. Intensity patterns did not deviate materially when compared with simulation of larger arrays of up to 600 nano-holes.

The EME simulation provided reflection spectra of FBGs (Fig. 2a, Fig. 4b-e, Fig. S1, Fig. S2 and Fig. S5) containing arrays of 600 or 1200 nano-holes that were spaced uniformly or π-shifted, respectively. A non-uniform spatial mesh was defined having the finest pitch for lateral dimension beginning at $\Delta y = \Delta z = 10$ nm inside the fibre core array, and gradually increasing to 400 nm in the cladding. A uniformly small pitch of $\Delta x = 6$ nm was maintained axially. A unit cell was defined around the $\Lambda = 1.072$ μm period of nano-holes and sandwiched by a pair of periodic boundary conditions. The program provided a fundamental single mode having mean field diameter (MFD) of ~10 μm and returned reflection spectra through the S-parameter function.

# Supplementary Information

Fig. S1. Simulated Spectra of Hollow-Array Fibre Bragg Gratings under varying Hole Diameter and Sensing Liquid

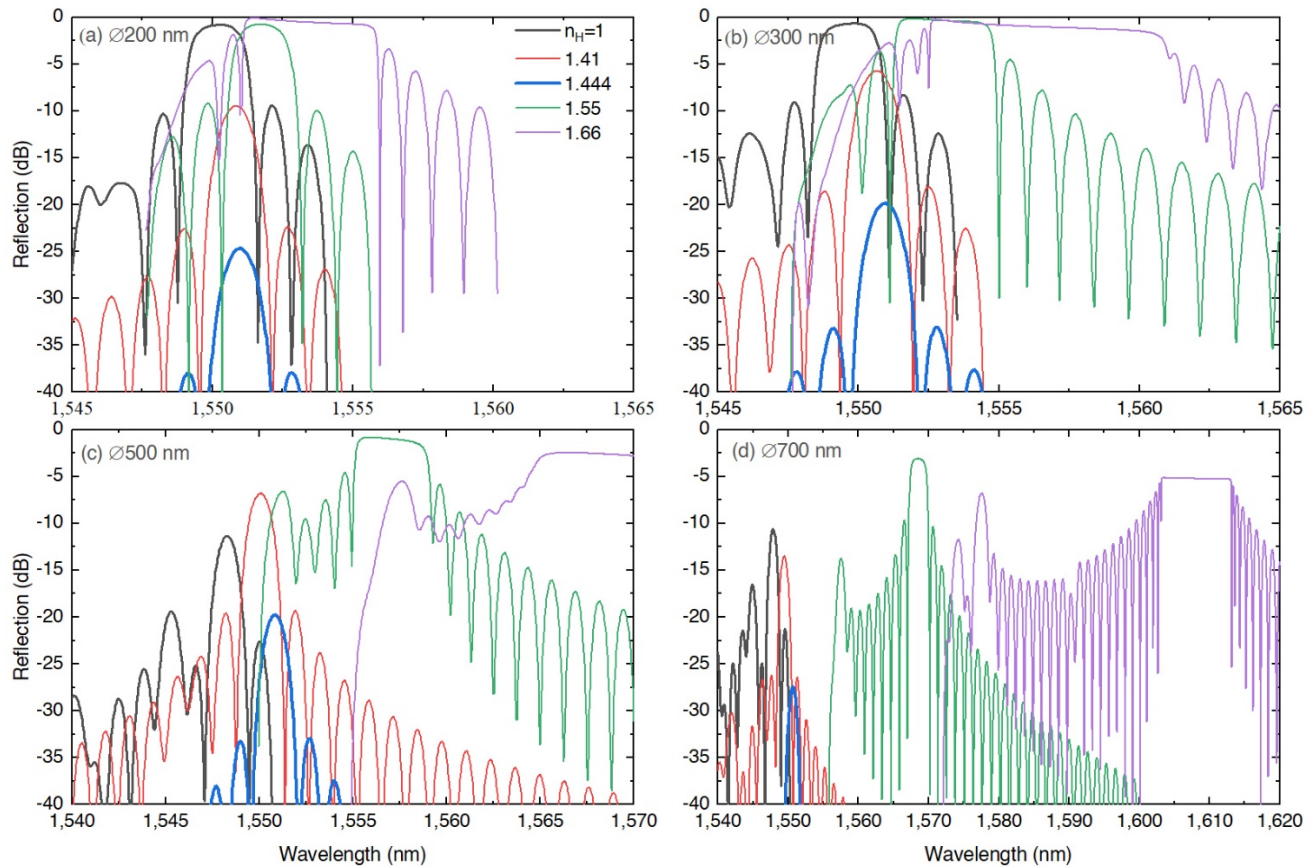

Fig. S1. Simulated reflection spectra (EME) generated for second order FBGs (Λ = 1072 nm) having 600 nano-holes spanning over 0.643 mm length. The relatively strong (up to -0.6 dB) and broad (1.2 nm to 11.1 nm) stopbands are noted to shift rapidly with wavelength as shown for the selected examples of air, solvents and oils ($n_H$ = 1.0 to 1.66) applied to fill the nano-holes. The further influence of nano-hole diameter (200 to 700 nm in (a) to (d)) demonstrates the potential of bandgap engineering in generating the strongest stopbands with the smaller range of nano-holes (200 to 300 nm). Alternatively, the largest wavelength shifts arise from the widest hole diameters (700 nm), suggesting a higher RI response. A wavelength shift of +62 nm is noted for at $n_H \cong 1.66$ with nano-hole diameter of 700 nm (d). By tuning the hole diameter, the simulated spectra could be closely matched with the experimentally recorded FBG spectra (Fig. 5f), providing an independent means from SEM in confirming the nano-diameters. Moreover, the close match of spectra confirms the precision of laser-driven filament explosion in providing highly reproducible holes of extra-ordinary aspect ratio on exceptionally close packing densities.

Fig. S2. Effect of Filament Number on the Reflection Strength and Bandwidth of Hollow-Array Fibre Bragg Gratings

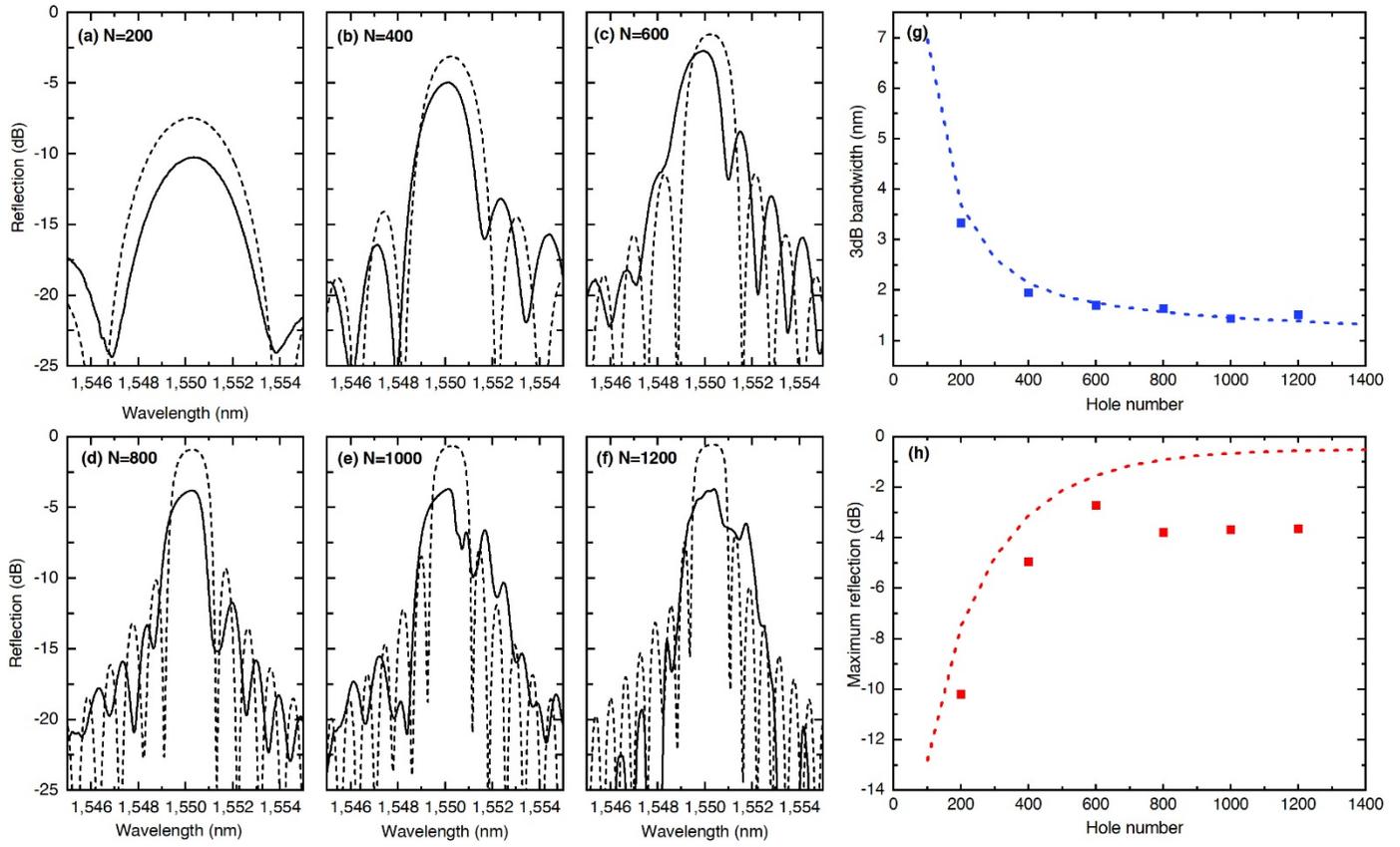

Fig. S2. The scaling of FBG reflection strength with the length of filament array is provided by the series of reflection spectra (a-f, solid lines) recorded from second-order gratings ($E_{pulse}$ = 4.5 μJ, $\Lambda$ = 1072 nm) with air filled holes ($n_H$ = 1.0). A comparison with EME simulated spectra (a-f, dashed line) shows close correspondence up to ~1000 holes, beyond which disorder in the filament sizes or relative positions in laboratory samples (solid lines) appear to accumulate in a broadening, skewing, and mixing of the main lobe peak with the adjacent side lobes. As the number of filaments increases from N = 200 to 1200, the observed bandwidth (g, squares) decreases from 7 nm to 1.5 nm in relative compliance with the EME simulation (g, blue dashed line). The peak reflection (h, squares) falls short by several dB of the simulated values (h, dashed line), increasing from -10 dB at 200 filaments to ~-2.5 dB for 600 filaments. Because gratings with reflection strengths greater than -2.7 dB could not be generated with the present apparatus for air-filled holes, all grating lengths were limited to a maximum of 1.286 mm (i.e., 1200 filaments).

Fig.S3. Tuning of Bragg Resonance in Hollow-Arrayed Fibre Gratings by Chemical Etching

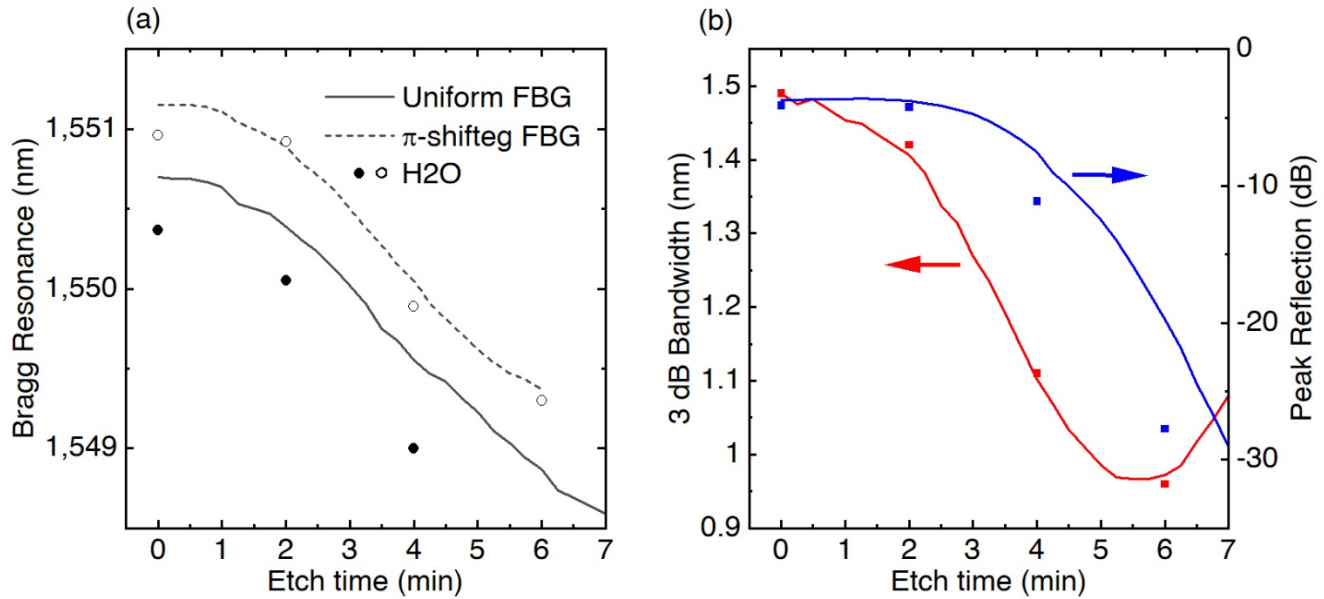

Fig. S3. Real-time recording of the influence on the Bragg grating resonance wavelength during chemical etching of second order nano-holed FBGs ($E_{pulse}$=4.5 µJ, $\Lambda$ = 1072 nm) with 5% diluted HF acid (a), comparing a uniform 600 filament array (solid line) with a π-shifted array of 1200 filaments (dashed line). Similar spectral shifts are noted for the uniform (solid circle) and π-shifted (open circle) gratings that were removed at different etching times of 0, 2, 4, and 6 min and filled with water. The moderately strong ~2 nm shift to shorter wavelength attests to an increasing diameter of the nano-holes when core waveguide glass has been displaced with a lower RI acid ($n_H$ =1.36). For the 600-filament array, the increasing hole diameter resulted in a dramatic fall-off of the peak reflectance from -5 dB to ~-30 dB over a 7 min etching time (b, blue line). In contrast, the Bragg linewidth (3dB) sharpened from 1.5 to 1 nm over a 6 min etching time (b, red line). The peak reflection (b, blue squares) and bandwidth (b, red squares) followed similar trends when the 600 filament FBG was removed after 0, 2, 4, and 6 min of chemical etching and filled with water. The dramatic drop in the peak reflectance to 6 min etching time arises from a hole diameter expanding above an optimal ~$\lambda/4n_{eff}$ diameter for high reflection beginning in the as-formed FBG (0 min). The rising linewidth and weakening FBG observed beyond 6 min of etching points to the present limit of chemical etching with the onset of nano-hole degradation as holes begin to merge into each other.

Fig. S4. Spectral Recordings of Fibre Bragg Stop-Bands Under Varying Hole Diameter and Sensing Liquid

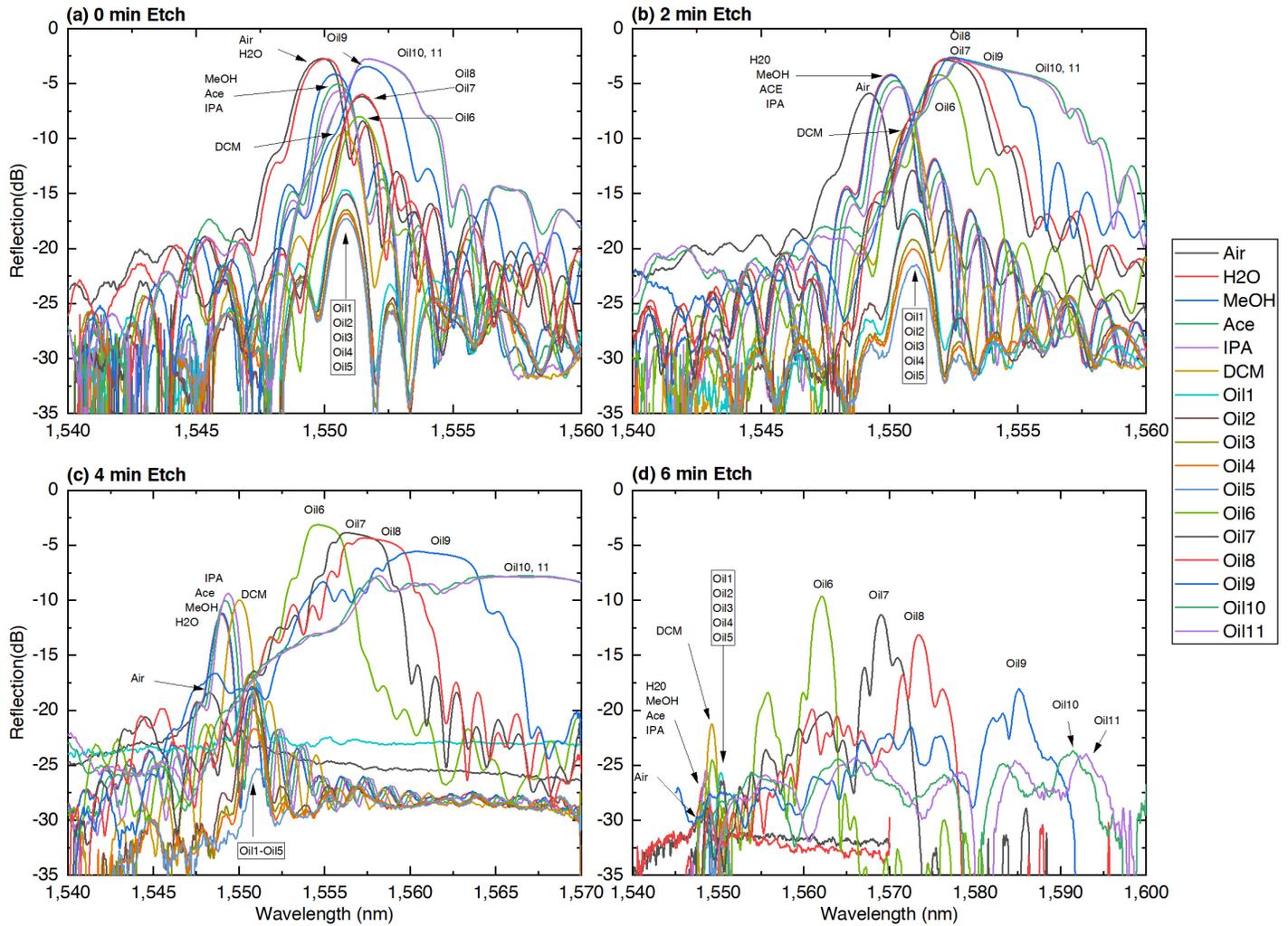

Fig. S4. The complete set of the reflection spectra recorded from FBGs having 600 nano-hole filaments ($E_{pulse}$ = 4.5 µJ, Λ = 1072 nm) and their widely varying response to varying RI liquids (colour coded and as labelled in (a)) after undergoing 0 (a), 2 (b), 4 (c) and 6 (d) minutes of chemical etching time. The data complement the representative samples of spectra provided in Fig. 4a and 5a-c. The central wavelength, linewidth and peak reflection of the Bragg resonance for each combination of liquid and chemical etching time were extracted from this data set and plotted in Fig. 5d, e, and f, respectively. The data show that the nano-holes with a smaller diameter such as the unetched case in (a) provide the strongest reflection resonances for a majority of the liquids evaluated, while the increasing larger hole diameters (b to d) provided an increasing Bragg wavelength shift for improving refractive index sensitivity, as shown in Fig. 5f (inset).

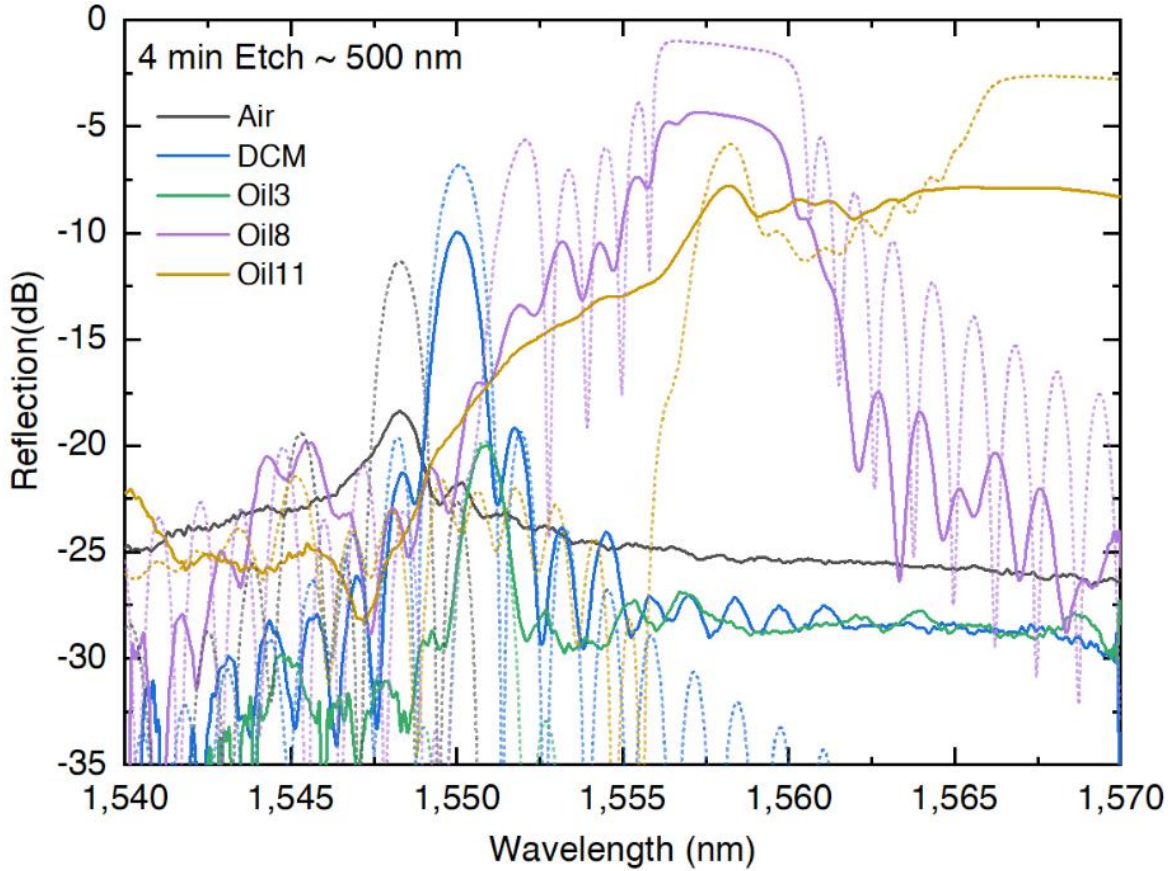

Fig. S5. Modelling the Bragg Stop-bands of Arrayed Nano-hole Fibre Gratings

Fig. S5. A comparison of experimentally recorded spectra (solid lines) with EME simulated spectra (dashed lines) for uniform FBGs ($E_{pulse}$ = 4.5 µJ, $\Lambda$ = 1072 nm, 600 holes) that have been chemically etched for 4 min. Only representative values of low to high RI values ($n_H$ = 1.0 to 1.67, see Table 1) have been adopted from Fig. 4c. A nano-hole diameter of 500 nm (±10 nm) provided the best alignment of simulated stopband and side-lobe positions to the observed spectra. Relatively symmetric stop bands and side lobes are noted in both of the simulation and experimental data for low to moderate refractive index cases ($n_H$ = 1 to 1.45). In contrast, the positive and high index contrast cases ($\Delta n > 1.562$) have developed an asymmetry that skewed stronger to shorter wavelength. With higher index contrast (Oil11), the short-wavelength band edge forms into a peak separated from a weakening but broadening stopband. An ambiguity in identifying the Bragg shift is thus reached for this case of large hole diameter and high positive contrast in refractive index, leading to a breaking trend of Bragg shifts in Fig. 5 (noted by the hollow triangle data points). With increasingly negative contrast in refractive index (Oil3, DCM, air), the stopbands weaken both in experimental recordings and simulation, that points to a nano-hole diameters of 500 nm approaching the anti-resonant reflection condition approximately at $\lambda/2n_H$ = 775 to 464 nm when $n_H$ = 1.0 to 1.67, respectively.

Fig. S6. Experimentally Recorded Spectra of π-shifted FBGs under varying Hole Diameter and Sensing Liquid

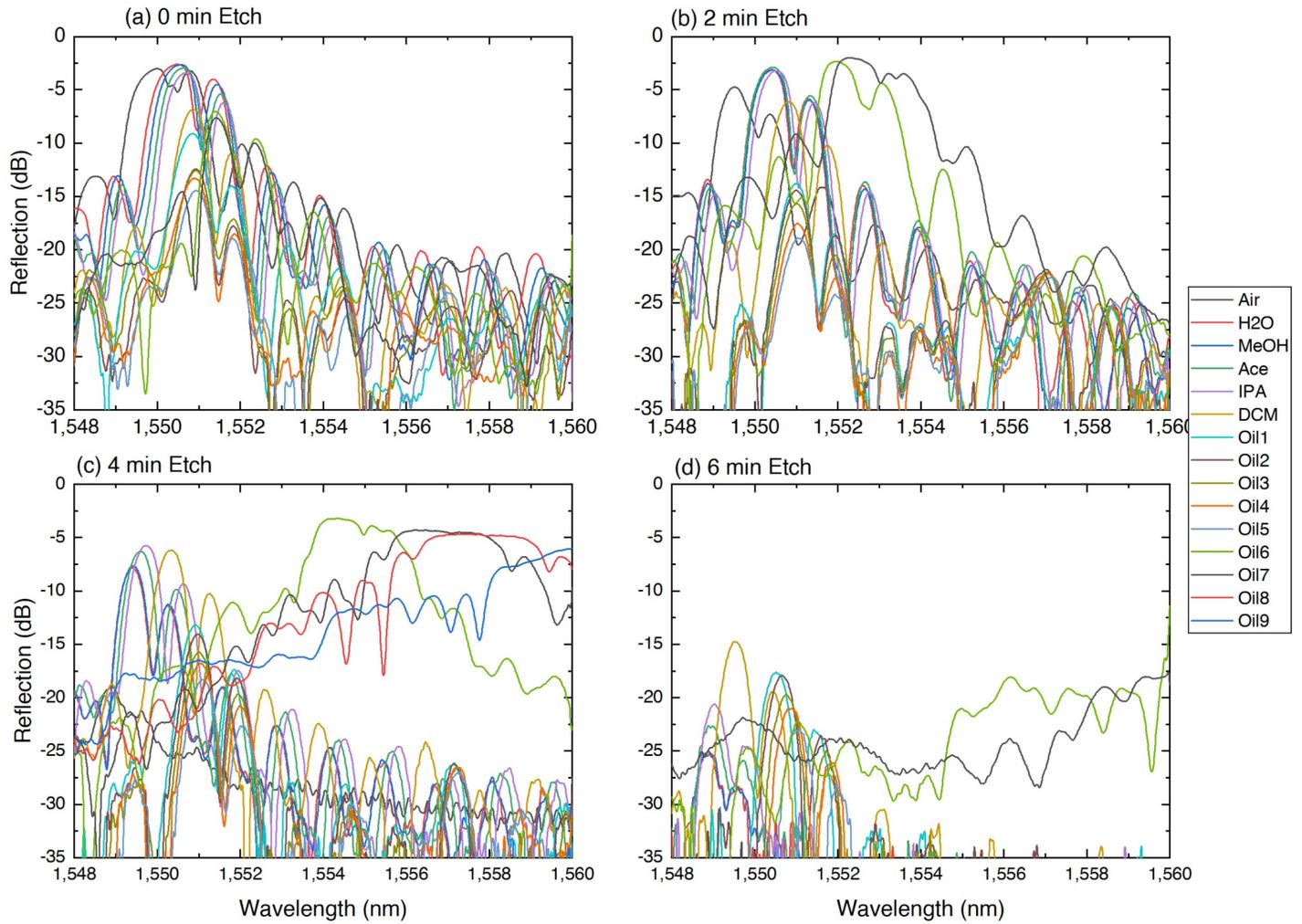

Fig. S6. The complete set of the reflection spectra recorded from π-shifted FBGs having 1200 nano-hole filaments ($E_{pulse}$ = 4.5 µJ, Λ = 1072 nm) and their widely varying response to liquids with varying RI values (colour coded and as labelled in (a)) after undergoing 0 (a), 2 (b), 4 (c) and 6 (d) minutes of chemical etching time. Representative samples of spectra were selected from here for presentation in Fig. 4b-e. The Bragg envelop for the present examples of π-shifted FBGs followed identical trends to those as reported for uniform FBGs in Fig. S4. Namely, the unetched nano-holes having smallest diameter (a) provided the strongest reflection resonances for a majority of the liquids evaluated, while an increasing hole diameter (b to d) provided an increasing Bragg wavelength shift for improving refractive index sensitivity, but with weakening overall reflection strength.

The π-shift resonance for laser-formed and chemically etched holes appear as narrow (50 to 370 pm) transmission windows that sharpened the sensing resolution of Bragg resonance. However, a birefringent broadening of the π-defect resonance plays out differently in the spectra according the refractive index contrast and hole diameter (See Fig. S7 for a full discussion). In the sequence of increasing hole diameter (a-d) for air filled holes (negative refractive index contrast, Δn = -0.45), the π-defect displays a moderately rising birefringence of $δλ_B$ = 210 to 370 pm for chemical etching times rising from 0 to 6 min. Birefringence broadening of $δλ_B ≅ 20$ pm is also strongly evident in select cases of positive refractive index contrast (i.e., Δn = 0.1 at 2 and 4 min etch for Oil6 and Oil7 in (b) or (c)) Otherwise, the birefringence is unresolved for a majority of the liquid solvents (1.31 < $n_H$ < 1.67) and etching time (0 – 6 min) presented here. EME modelling provided further insights into the broadening and birefringence splitting effects as discussed in Fig. S7.

Fig. S7. Spectral Resolution and Birefringence Response of π-shifted FBGs Observed under varying Hole Diameter and Sensing Liquid

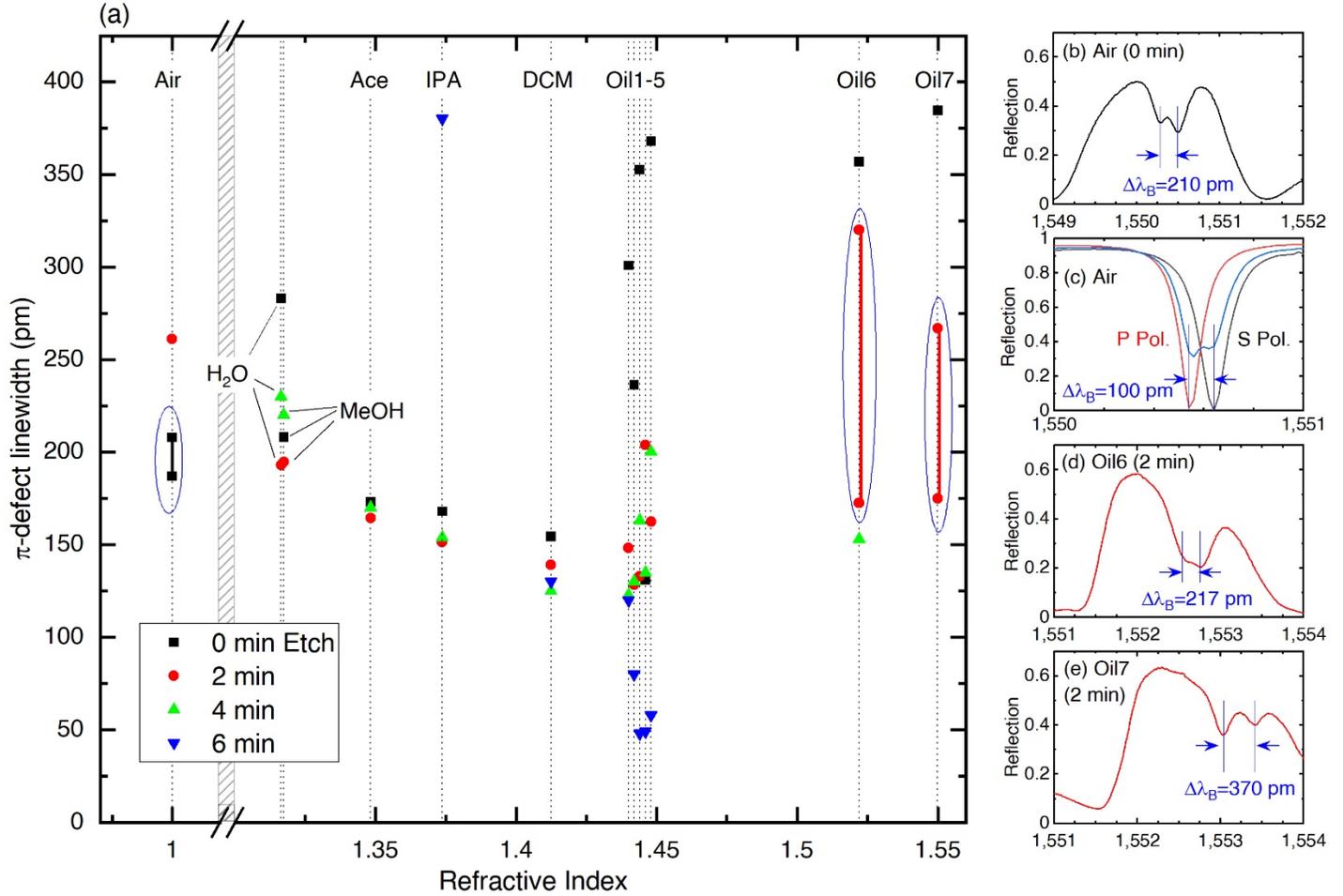

Fig. S7. The spectral recordings from π-shifted FBGs (Fig. S6; 1200 nano-holes, $E_{pulse}$ = 4.5 µJ, Λ = 1072 nm) revealed narrow defect linewidths varying from 50 to 370 pm for unpolarized light as plotted (a) for chemical etching times of 0 (black square), 2 (red circle), 4 (green triangle) and 6 (blue triangle) minutes. Centre wavelengths and 3 dB linewidths were obtained by Gaussian line shaping representations of the π-defect (see Methods). In this way, centre wavelengths and linewidths of the π-defects were followed with ±10 pm and ±5 pm precision, respectively. The sharpest π-defect resonances of 50 to 120 pm (a) were provided by holes filled with solvents having RI values ($n_H$= 1.440 to 1.448) slightly below that of the core waveguide. In these weaker stopbands, the π-defect narrowed to a minimum 50 pm linewidth as noted for the case of largest hole diameter, inferred to be around 700 nm (Fig. 5f) for the 6 min etching time. The EME modelling predicted a 135 pm linewidth with a birefringent wavelength splitting of 100 pm that closely matched with the observations. Otherwise, higher refractive index contrast generated much broader defect lines of up to 320 pm.

A large component of the π-defect broadening can be attributed to birefringent responses intrinsic to the nano-hole geometry and less so to radial stresses induced by the laser micro-explosion. This birefringence splitting was resolved in only a few of the recorded reflection spectra (Fig. S6) which have been reproduced in (b) for an air-filled, non-etched FBG (0 min) with Δn = -0.45 contrast, and in (d) and (e) for a 2 min etched FBG filled with Oil6 (Δn = +0.072) and Oil7 (Δn = +0.100), respectively. Examples of EME simulated spectra (c) for P (red line) and S (black line) polarization show a close alignment in centre wavelength positions with the recorded unpolarized light spectrum (blue line). The simulation provided a wavelength birefringence of $δλ_B = λ_S − λ_P$ = 100 pm that underestimates the observed $δλ_B$ = 210 pm as determined by Gaussian line fitting (b). Wavelength birefringence of $δλ_B$ = 217 and 370 pm as observed for Oil6 and Oil7, respectively, could not be definitively matched with simulated spectra. Since only stress-free nano-holes were simulated in the EME, the radial stresses expected around the nano-hole wall [49] as well as surface tensions effects may play underlying roles.

A narrowing of the π-defect resonances is an objective to further sharpen control of the FBG design that entails improving the nano-hole morphology and alignment positioning in the fibre core.

Media: Optical Microscope video recordings of Isopropanol Wetting and Evaporation of the Nano-hole array grating in Optical Fibre

Optical microscope video recordings (400× magnification) of the optical fibre cross-section (125 µm diameter) have resolved the nano-hole array formed with 200 nm diameter and 1072 nm period in standard optical communication fibre (Corning, SMF-28). The recordings capture the filling and evaporation of isopropanol into and from the nano-hole array, with the meniscus confirming that the nano-holes are fully open from cladding-to-cladding without showing hole-to-hole break-through.

**Video 1:** Filling of isopropanol into nano-holes passing through the silica cladding and guiding core of an optical fibre.

**Video 2:** Evaporation of isopropanol from nano-holes passing through the silica cladding and guiding core of an optical fibre.

Link:

https://dataverse.scholarsportal.info/dataset.xhtml?persistentId=doi%3A10.5683%2FSP2%2FY7T3WW&fbclid=IwAR0kDgmRb_N96uZNrkQl1iNQVo3SfIMtBbZ7VCaDb7abjItlwGxYu-dl06k